\documentclass[%
 reprint,
 amsmath,amssymb,
 aps,
floatfix,
]{revtex4-2}

\usepackage[dvipdfmx]{graphicx}
\usepackage{dcolumn}
\usepackage{bm}
\usepackage{braket}
\usepackage{natbib}
\usepackage{xcolor}
\usepackage{mathtools}
\begin{document}

\preprint{APS/123-QED}

\title{Quantum Kernel t-Distributed Stochastic Neighbor Embedding}

\author{Yoshiaki Kawase}
\email{ykawase@g.ecc.u-tokyo.ac.jp}
\affiliation{%
Graduate School of Information Science and Technology, The University of Tokyo\\
7-3-1 Hongo, Bunkyo-ku, Tokyo 113-8656, Japan
}%
\affiliation{%
Graduate School of Engineering Science, Osaka University\\
1-3 Machikaneyama, Toyonaka, Osaka 560-8531, Japan
}

\author{Kosuke Mitarai}%
 \email{mitarai@qc.ee.es.osaka-u.ac.jp}
\affiliation{%
Graduate School of Engineering Science, Osaka University\\
1-3 Machikaneyama, Toyonaka, Osaka 560-8531, Japan
}
\affiliation{%
Center for Quantum Information and Quantum Biology, Osaka University 560-0043, Japan.
}%

\author{Keisuke Fujii}
\email{fujii@qc.ee.es.osaka-u.ac.jp} 
\affiliation{%
Graduate School of Engineering Science, Osaka University\\
1-3 Machikaneyama, Toyonaka, Osaka 560-8531, Japan
}
\affiliation{
Center for Quantum Information and Quantum Biology, Osaka University 560-0043, Japan.
}%
\affiliation{
RIKEN Center for Quantum Computing, Wako Saitama 351-0198, Japan.
}

\date{\today}

\begin{abstract}
Data visualization is important in understanding the characteristics of data that are difficult to see directly. 
It is used to visualize loss landscapes and optimization trajectories to analyze optimization performance. 
Popular optimization analysis is performed by visualizing a loss landscape around the reached local or global minimum using principal component analysis. 
However, this visualization depends on the variational parameters of a quantum circuit rather than quantum states,
which makes it difficult to understand the mechanism of optimization process through the property of quantum states. 
Here, we propose a quantum data visualization method using quantum kernels, 
which enables us to offer fast and highly accurate visualization of quantum states. 
In our numerical experiments, we visualize hand-written digits dataset and apply $k$-nearest neighbor algorithm to the low-dimensional data to quantitatively evaluate our proposed method compared with a classical kernel method. 
As a result, our proposed method achieves comparable accuracy to the state-of-the-art classical kernel method, meaning that 
the proposed visualization method based on quantum machine learning does not degrade the separability of the input higher dimensional data.
Furthermore, we visualize the optimization trajectories of finding the ground states of transverse field Ising model and successfully find the trajectory characteristics. 
Since quantum states are higher dimensional objects that can only be seen via observables, 
our visualization method, which inherits the similarity of quantum data, would be useful in understanding the behavior of quantum circuits and algorithms.
\end{abstract}

\keywords{data visualization, quantum kernel, quantum machine learning, visualizing optimization trajectories}
\maketitle


\section{\label{sec:introduction}Introduction}
Data visualization is a popular and important application of dimensionality reduction, 
such as principal component analysis (PCA), autoencoder \cite{hinton2006reducing}, and variational autoencoder \cite{kingma2013auto}. 
Data visualization is performed by mapping high-dimensional data into two-dimensional space. 
The visualized data are useful in finding trends and understanding the characteristics of data 
that are difficult to see directly from the raw high-dimensional data. 
PCA and t-distributed stochastic neighbor embedding (t-SNE) \cite{van2008visualizing} are popular choices 
for data visualization in classical machine learning. 
For example, they are used to visualize loss landscapes \cite{li2018visualizing} and optimization processes \cite{erhan2010does,fort2019deep} to gain intuitive insights into the complex optimization processes. 

In quantum computing, there is a similar demand to obtain an intuitive insight 
into the optimization of quantum circuit architecture design or variational quantum algorithms. 
More concretely, variational quantum algorithms \cite{cerezo2021variational}, 
such as variational quantum eigensolver \cite{peruzzo2014variational,kandala2017hardware}, quantum approximate optimization algorithm \cite{farhi2014quantum}, 
and quantum neural networks \cite{mitarai2018quantum,farhi2018classification,schuld2014quest},
are promising routes to use noisy intermediate-scale quantum devices for practically important problems.
However, it is difficult to design a parameterized quantum circuit for a variational ansatz 
because different tasks require different expressibility
\cite{sim2019expressibility,nakaji2021expressibility}.
Furthermore, once a parameterized quantum circuit is fixed, 
there exist challenges in parameter optimization, such as vanishing gradient problem \cite{cerezo2021cost,mcclean2018barren,wang2021noise} and being trapped in local minima \cite{bittel2021training}.
Quantum data visualization \cite{rudolph2021orqviz,kawase2022parametric}
can be used to address these issues in variational quantum algorithms. 
Indeed, 
visualizing loss landscapes and optimization trajectories by principal component analysis
have been performed in quantum machine learning \cite{rudolph2021orqviz}. 
However, this approach only visualizes changes in variational parameters 
and does not reflect direct information about quantum states during optimization process.

To visualize quantum states, 
we have proposed a variational quantum algorithm to visualize classical and quantum data using a quantum neural network \cite{kawase2022parametric} based on t-SNE.
Specifically, we map quantum states representing high-dimensional data onto a two-dimensional plane 
by optimizing a parameterized quantum circuit 
to preserve the similarities between the high and low-dimensional data. 
However, this method requires too many accesses to a quantum computer 
to optimize the design and parameters of a quantum circuit applied to quantum states. 
The algorithm sometimes does not work in a practical timescale, even on a noiseless simulator using a few qubits. 
Additionally, we must design a quantum circuit to avoid barren plateaus and local minima. 
These issues make it difficult to visualize quantum states. 

To resolve these issues and realize faster and easier quantum data visualization, 
we here propose a visualization method based on a quantum kernel method \cite{havlivcek2019supervised,schuld2019quantum}, 
which uses inner products between quantum states representing data
to perform machine learning tasks, e.g., regression or classification. 
Specifically, we employ quantum kernels instead of classical kernels in 
kernel t-SNE \cite{Gisbrecht2012OutofsampleKE,bunte2012general,gisbrecht2015parametric}, which is a kernel-based extension of the original t-SNE. 
Since the kernel t-SNE only uses kernel values between data during its visualization process, we can fix the number of quantum circuit execution before starting the algorithm, as opposed to our previous approach which iteratively optimizes a circuit for better visualization.
Consequently, our proposed method requires fewer accesses to a quantum computer in most cases than our previous quantum neural network approach \cite{kawase2022parametric}.
Moreover, because a parameterized quantum circuit applied to quantum states is not required,
we can visualize quantum states more directly using our proposed method than our previous one \cite{kawase2022parametric}.

To verify our proposed method in terms of visualization performance, 
we first visualize quantum states generated through a quantum feature map
from hand-written digits data set \cite{Dua:2019} in a two-dimensional plane.
To quantitatively evaluate our model compared with our previous and classical machine learning models, 
we apply the $k$-nearest neighbor algorithm to the low-dimensional data 
and compare the accuracy of the classification.
We confirm the prediction by our proposed method is more accurate 
than by our previous quantum neural network approach \cite{kawase2022parametric}. 
Moreover, our method achieves the accuracy mostly equivalent to the kernel t-SNE using Gaussian kernels.


Secondly, we use the method for visualizing quantum states generated during the optimization process of the variational quantum eigensolver (VQE) \cite{peruzzo2014variational, cerezo2021variational} algorithm applied to transverse field Ising models, which we could not perform by the previous method \cite{kawase2022parametric}. 
This visualization reveals the effect of parameter initialization and that a part of a trajectory is shared with other trajectories starting from different initial points. 
We believe that visualizing the optimization process of variational quantum algorithms will be helpful in improving optimization strategies.


The rest of this paper is organized as follows.
In Sec.~\ref{sec:preliminary}, we describe the data visualization method of t-SNE and 
the extension of t-SNE. 
In Sec.~\ref{sec:proposed_method}, we explain our proposed method to visualize data by quantum kernels. 
In Sec.~\ref{sec:numerical_experiments}, 
to verify our proposed method, 
we visualize a hand-written digits data set embedded in quantum states via a quantum feature map, 
and compare the prediction accuracy by applying $k$-nearest neighbor algorithm to the low-dimensional data. 
Additionally, 
we visualize the optimization trajectories of a variational quantum algorithm. 
We describe our conclusion and future work in Sec.~\ref{sec:conclusion}.

\section{\label{sec:preliminary}Preliminary}
In this section, we give a rough overview of existing data visualization methods. 
First, we introduce t-SNE in Sec.~\ref{subsec:t-sne}. 
Then, we explain an extension of t-SNE called parametric t-SNE, which uses neural networks, quantum neural networks, and kernel methods
in Secs.~\ref{subsec:parametric_tsne}, \ref{subsec:qnn_parametric_tsne}, and \ref{subsec:kernel_tsne}.

\subsection{t-SNE} \label{subsec:t-sne}
t-SNE \cite{Maaten2008VisualizingDU} is a non-parametric unsupervised learning method
to visualize data. 
More concretely, t-SNE maps high-dimensional data $\bm{x}_i$ to lower dimensional data $\bm{y}_i$
while preserving the similarities between high- and low-dimensional data. 
The high-dimensional data similarities $p_{ij}$ between data $\bm{x}_i$ and $\bm{x}_j$ are defined using a Gaussian distribution:
\begin{align} \label{eqn:similarity_high_dim}
p_{j|i} &= \frac{ \exp{( -\|\bm{x}_i-\bm{x}_j\|^2/2\sigma_i^2 )} }{ \sum_{k \neq i} \exp{( -\| \bm{x}_i-\bm{x}_k\|^2/2\sigma_i^2 )} }, \nonumber \\
p_{ij} &= \frac{p_{j|i}+p_{i|j}}{2N}, \\
p_{ii} &= 0 \nonumber ,
\end{align}
where $N$ denotes the number of data, 
and $\sigma_i$ is a parameter determined by making a quantity called perplexity $\mbox{Perp}_i=2^{-\sum_j p_{j|i} \log_2{p_{j|i}}}$ equal to a user-specified value. 
The low-dimensional data similarities are defined using t-distribution with one degree of freedom: 
\begin{align} \label{eqn:similarity_low_dim}
q_{ij} &= \frac{ (1+\|\bm{y}_i-\bm{y}_j\|^2)^{-1} }{ \sum_k \sum_{l \neq k} (1+\|\bm{y}_k-\bm{y}_l\|^2)^{-1}}, \\
q_{ii} &= 0. \nonumber
\end{align}

In t-SNE, the initial positions of low-dimensional data are determined at random. 
The low-dimensional data are then iteratively moved to minimize the cost function $C$
defined by Kullback-Leibler(KL) divergence between the similarities of data distributions
in the high- and low- dimensional spaces: 
\begin{equation} \label{eqn:costf}
C = KL(P \| Q)=\sum_i \sum_j p_{ij} \log{ \frac{p_{ij}}{q_{ij}} }.
\end{equation}
After the optimization, we can visualize data by plotting the resulting low-dimensional data. 
A caveat of t-SNE is that we cannot add data after the visualization; to add data to the visualization, we must redo the whole process from the beginning. 
This short-coming is resolved by modified versions of t-SNE which are discussed in the following subsections. 

\subsection{\label{subsec:parametric_tsne} Parametric t-SNE}
Parametric t-SNE \cite{van2009learning} is the variant of t-SNE
that can be applied to unseen data.
It explicitly constructs a map from a high-dimensional space to a low-dimensional space by a neural network.
Let us denote weights of a neural network as $\bm{w}$.
A neural network $\bm{y}=f(\bm{x}|\bm{w})$ is trained to minimize the cost function $C$ in Eq.~\eqref{eqn:costf}.
The trained network allows us to visualize data, including unseen ones.
In numerical experiments, we use parametric t-SNE 
to compare the accuracy of our proposed method.

\subsection{\label{subsec:qnn_parametric_tsne} Parametric t-SNE with a quantum neural network}
We have proposed to use a quantum neural network instead of a neural network in parametric t-SNE to visualize both classical and quantum data \cite{kawase2022parametric}. 
Specifically, we apply a parametrized quantum circuit to a set of quantum states $\{\ket{\psi_i}\}$ to visualize them.
The states $\{\ket{\psi_i}\}$ can either be intrinsically quantum, e.g. quantum states provided from physical experiments or from other quantum circuits, or those that encode classical data $\{\bm{x}_i\}$ through some quantum feature map. 
In this method, the high-dimensional data similarities are defined as
\begin{align} \label{eqn:similarity_high_dim_redef}
    p_{j|i} &= \frac{ \exp{( -d( \bm{x}_i, \bm{x}_j )^2/2\sigma_i^2 )} }{ \sum_{k \neq i} \exp{( -d( \bm{x}_i,  \bm{x}_k )^2/2\sigma_i^2 )} }, \nonumber \\
    p_{ij} &= \frac{p_{j|i}+p_{i|j}}{2N}, \\
    p_{ii} &= 0, \nonumber 
\end{align}
where
\begin{equation} \label{eqn:distance}
\begin{cases}
d( \bm{x}_i , \bm{x}_j )^2 = \|\bm{x}_i-\bm{x}_j\|^2 & \mbox{(for classical data)}\\
d( \ket{\psi_i} , \ket{\psi_j} )^2 = 1-|\langle \psi_i | \psi_j \rangle |^2 & \mbox{(for quantum data)}
\end{cases},
\end{equation}
and $\|\cdot\|$ is Euclidean norm.
On the other hand, the low-dimensional data $y_{\mu}$ is given by
\begin{equation*}
    y_{\mu}(\bm{x}_i, \bm{\theta}) = \beta_\mu \bra{0}U^\dagger(\bm{x}_i,\bm{\theta}) O_\mu U(\bm{x}_i,\bm{\theta})\ket{0}
\end{equation*}
for classical data, and 
\begin{equation*}
    y_{\mu}(\ket{\psi_i}, \bm{\theta}) = \beta_\mu \bra{\psi_i}U^\dagger(\bm{\theta}) O_\mu U(\bm{\theta})\ket{\psi_i}
\end{equation*}
for quantum data. 
Note that $U(\bm{x}_i,\bm{\theta})$ or $U(\bm{\theta})$ is a parameterized quantum circuit, 
$\{O_\mu\}_{\mu=1}^{d}$ are observables, 
and $d$ is the dimension of low-dimensional space. 
Also, note that $\beta_\mu$ is used as a trainable parameter instead of a hyper-parameter in Ref.~\cite{kawase2022parametric}
to increase expressibility in our numerical experiments.


\subsection{\label{subsec:kernel_tsne}Kernel t-SNE} 
Kernel t-SNE is the variant of t-SNE that can be applied to unseen data \cite{Gisbrecht2012OutofsampleKE,bunte2012general}.
Given a training dataset $\{\bm{x}_i\}$, the task here is to visualize a high-dimensional vector $\bm{x}$ on a $d$-dimensional space along with $\{\bm{x}_i\}$.
The kernel t-SNE maps $\bm{x}$ to $d$-dimensional data (low-dimensional data) $\bm{y}$ as
\begin{equation} \label{eqn:kernel_tsne}
    \bm{y}( \bm{x} )=\sum_{i=1}^N \bm{\alpha}_i k( \bm{x}_i, \bm{x} ), 
\end{equation}
where $N$ is the number of training data, $k$ is a kernel function, 
and $\bm{\alpha}_i \in \mathbb{R}^d$ are trainable parameters. 
In classical machine learning,  
a Gaussian kernel $k(\bm{x_i},\bm{x})=\exp{(-\|\bm{x_i}-\bm{x}\|^2/(2{\sigma^\prime}^2))}$ is often employed as the kernel function. 
The parameters $\bm{\alpha}_i$ are determined 
by minimizing the t-SNE cost function \cite{Gisbrecht2012OutofsampleKE,bunte2012general}. 
This method is expected to accelerate the mapping of all data points 
through learning the map of t-SNE consisting of the subset. 
In this paper, we use this method to accelerate the visualization in terms of the number of access to a quantum computer. 
Specifically, we propose to use quantum kernels defined by inner products of quantum states.
This allows us to visualize them
without optimizing a parametrized quantum circuit in contrast to our previous work \cite{kawase2022parametric}. 

\section{\label{sec:proposed_method}Quantum Kernel \lowercase{t}-SNE}
Here, we describe our proposal, that is, a visualization method using quantum kernels 
and its advantages over the previous one \cite{kawase2022parametric}. 

\begin{figure*}
    \centering
	\includegraphics[width=0.9\linewidth]{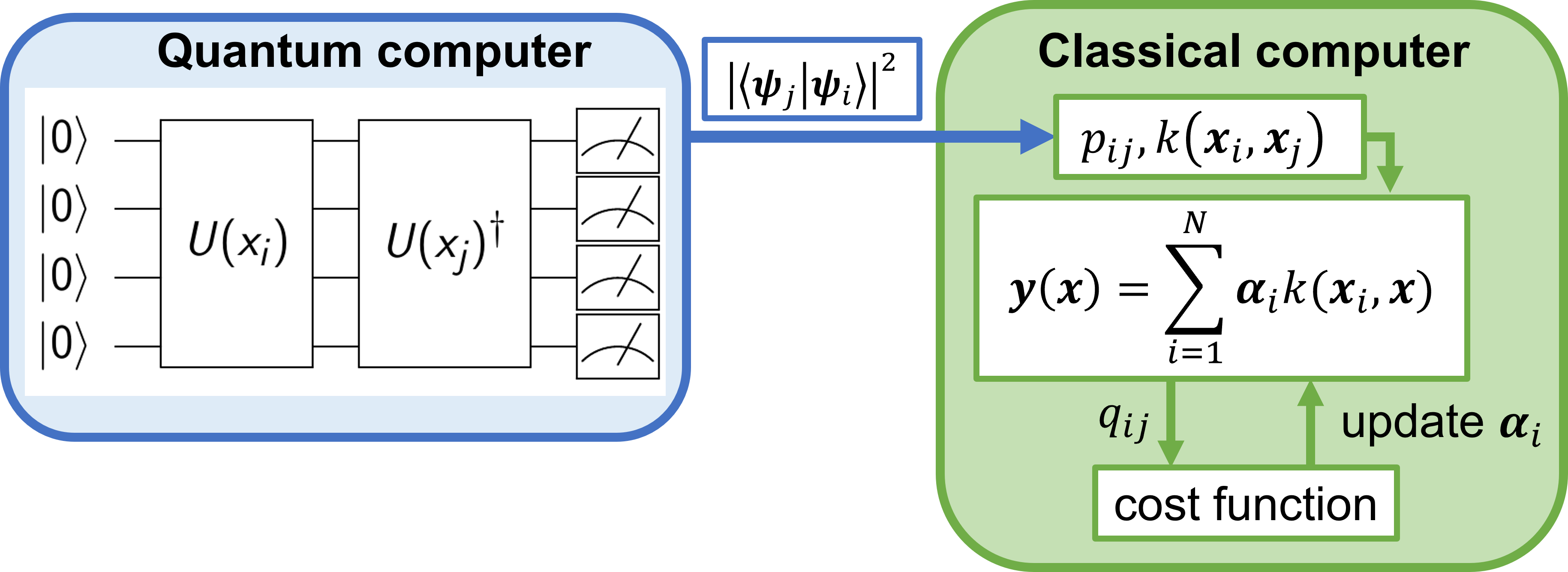}
	\caption
        {
        The outline of our proposed method. 
        After the fidelities between quantum states representing data are evaluated for all data pairs on a quantum computer, 
        the rest of the optimization process is performed entirely on a classical computer. 
        The optimization minimizes the cost function by updating $\bm{\alpha}_i$. 
        After the optimization, we visualize data by plotting the low-dimensional data $\bm{y}$.
        }
	\label{fig:flow_chart}
\end{figure*}

\subsection{\label{subsec:alg_quantum_kernel_tsne} Algorithm} 
Before we describe our proposed method, 
we first refer to our notation.
We are interested in visualizing a quantum state $\ket{\psi}$ along with a set of quantum states $\{\ket{\psi_i}\}_{i=1}^N$ provided as a training dataset.
As in Sec.~\ref{subsec:qnn_parametric_tsne}, the states $\{\ket{\psi_i}\}$ and $\ket{\psi}$ can either be intrinsically quantum, e.g. quantum states provided from physical experiments or from other quantum circuits, or those that encode classical data $\{\bm{x}_i\}$ and $\bm{x}$ through some quantum feature map.

The outline of our proposed method is shown in Fig.~\ref{fig:flow_chart}.  
The procedure of our proposed method is described as follows:
\begin{enumerate}
    \item Calculate the high-dimensional data similarities $p_{ij}$ for all data pairs by Eqs.~\eqref{eqn:similarity_high_dim_redef} and \eqref{eqn:distance}.
    \item Calculate the low-dimensional data $\bm{y}$ through Eq.~\eqref{eqn:kernel_tsne} with $k(\ket{\psi_i}, \ket{\psi}) = |\langle \psi_i | \psi\rangle |^2$.
    \item Calculate the low-dimensional data similarities $q_{ij}$ and the cost function $C$ defined as Eq.~\eqref{eqn:costf}.
    \item Update $\bm{\alpha}_i$ to minimize the cost function $C$.
\end{enumerate}
After the optimization, we can visualize data by plotting the low-dimensional data $\bm{y}$ using the optimized $\{\bm{\alpha}_i\}$. 

\subsection{Advantage over the previous method}
In this section, 
we explain the advantages of our proposed method compared to our previous approach  \cite{kawase2022parametric}. 
Specifically, we compare the number of accesses to a quantum computer. 
Let us define $p$ as the number of parameters in a quantum circuit 
and $m$ as the number of iterations required to optimize the parameters within a circuit in our previous approach. 

First, we compare the number of accesses to a quantum computer when visualizing classical data.
In the present method, 
we need to evaluate $O(N^2)$ fidelities to calculate the kernel function during training. 
For our previous approach \cite{kawase2022parametric}, 
we need to use $O(Np)$ accesses for each iteration
to evaluate the cost function and its gradient. 
So, $O(Npm)$ accesses are required for the whole optimization. 
The present method is therefore advantageous roughly when $mp>N$ in terms of the number of accesses to a quantum computer compared to the previous work. 
This condition is satisfied in a common situation. 
For example, 
in Ref.~\cite{du2021learnability}, 
they use a quantum neural network with $15$ or $60$ parameters and optimize the parameters with $400$ iterations 
to classify $360$ data into two labels. 
If we visualize the data, 
the present method performs visualization more efficiently than the previous one
under the condition that the latter requires roughly the same number of iterations for the optimization of the quantum neural network as in Ref.~\cite{du2021learnability}. 

Next, we compare the number of accesses when visualizing quantum data. 
We need $O(N^2)$ accesses in the present method. 
On the other hand, 
the previous one \cite{kawase2022parametric} 
needs $O(N^2)$ accesses to evaluate fidelities 
to calculate high-dimensional data similarities. 
Additionally, it needs $O(Np)$ accesses for each iteration
to evaluate the cost function and its gradient. 
In total, $O(N^2)+O(Nmp)$ accesses are required for the whole optimization. 
Therefore, our proposed method is more efficient for the visualization of quantum data than our previous work \cite{kawase2022parametric} without any assumptions in contrast to the case of classical data. 
We summarize the results in Table~\ref{tbl:n_qc_access}.

\begin{table}
\caption{
This table shows the number of accesses to a quantum computer. 
In the table, $N$, $m$, and $p$ denote the number of data, iterations required in the optimization of parameters in a quantum circuit, and parameters in a parametrized quantum circuit, respectively. 
\label{tbl:n_qc_access}}
\begin{ruledtabular}
\begin{tabular}{ccc}
Data type & classical  & quantum  \\ 
\hline
Parametric t-SNE with QNN \cite{kawase2022parametric} & $O(Nmp)$ &  $O(N^2)+O(Nmp)$ \\  
This work & $O(N^2)$ & $O(N^2)$ \\ 
\end{tabular}
\end{ruledtabular}
\end{table}

\section{\label{sec:numerical_experiments}Numerical Experiments}
In this section, first,
we perform numerical experiments to validate our proposed method 
by visualizing classical data. 
We apply the $k$-nearest neighbor algorithm to low-dimensional data 
to quantitatively evaluate the visualization performance. 
Then, 
we present a visualization of optimization trajectories in a variational quantum eigensolver. 
From the visualization, we explore the characteristics of ansatz's optimization trajectories. 
We use Qulacs \cite{suzuki2021qulacs} to classically simulate quantum circuits in our numerical experiments. 

\subsection{\label{subsec:numerical_experiments_classical}Visualizing classical data}
\begin{figure*}
    \centering
	\includegraphics[width=0.9\linewidth]{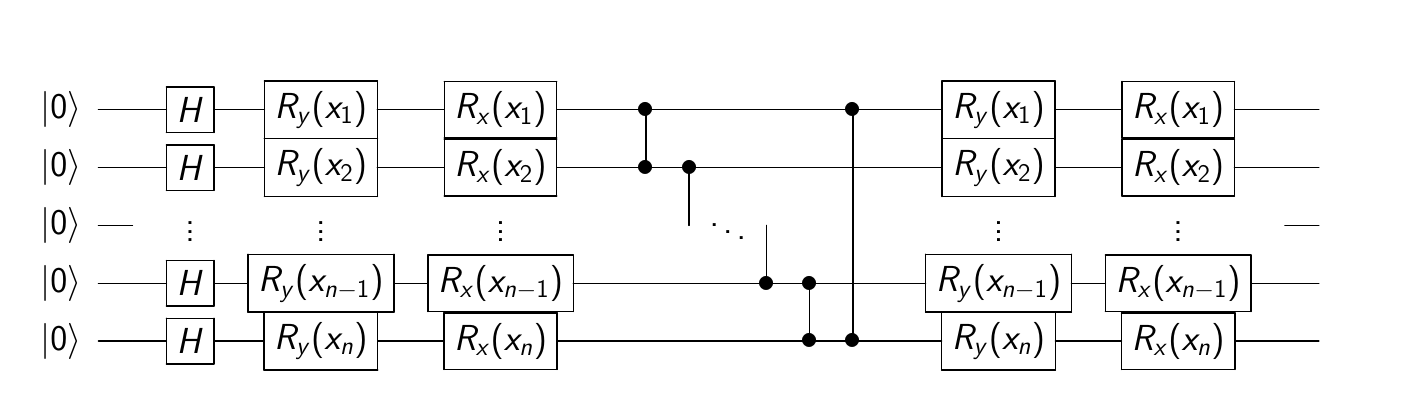}
	\caption{
        The encoding circuit for the hand-written digits data set. 
        In the figure, $x_i$ is the $i$th principal component of a data $\bm{x}$.
	}
	\label{fig:qc_reupload}
\end{figure*}

\begin{figure*}
    \centering
	\includegraphics[width=0.98\linewidth]{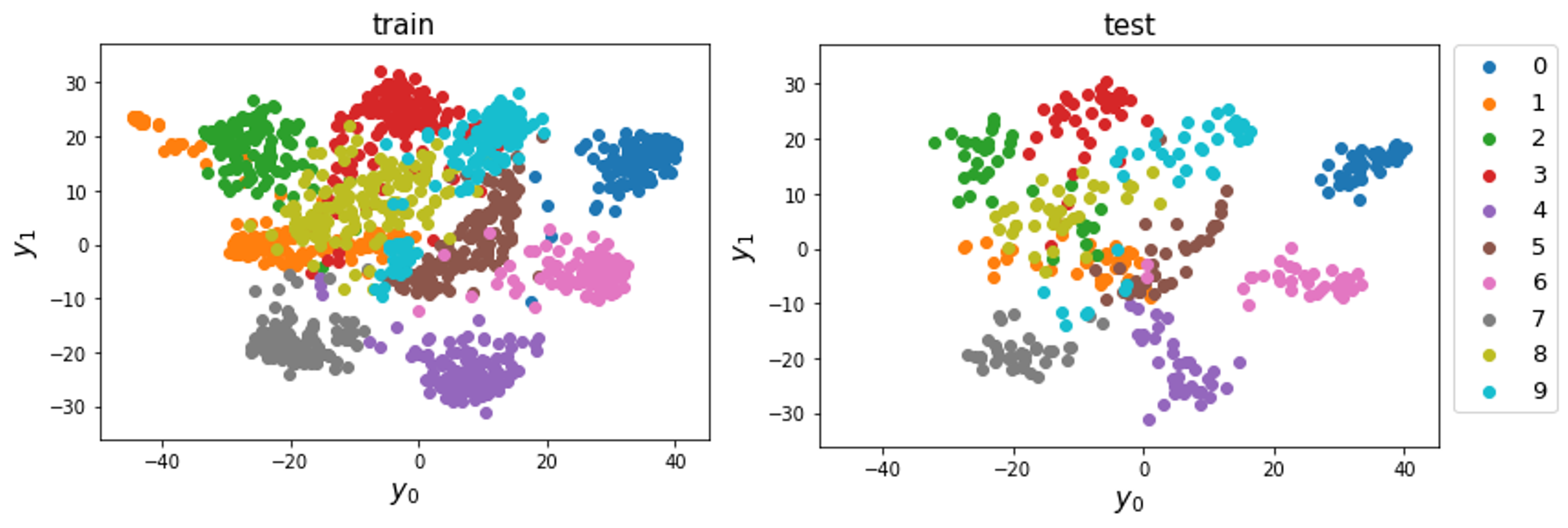}
	\caption{
        These figures visualize the hand-written digits data set,  
        where we split $80\%$ of the given data as training data 
        and the remaining $20\%$ as test data.
        The left figure visualizes the training data 
        and the right one visualizes the test data. 
        }
	\label{fig:vis2d}
\end{figure*}

Here, we visualize hand-written digits data set provided in scikit-learn \cite{Dua:2019} 
as an example of visualizing classical data. 
Specifically, we split $80\%$ of the given data as training data 
and the remaining $20\%$ as test data.
We employ the quantum feature map 
defined by a quantum circuit described in Fig.~\ref{fig:qc_reupload} with $12$ qubits. 
As a preprocess, 
we reduce the dimension of the data to $12$
by PCA.
Additionally, we normalize each principal component from $0$ to $\pi/2$ so that the data can be input as the angles of quantum gates. 
As shown in Fig.~\ref{fig:qc_reupload}, we input the $i$th principal component $x_i$ as angles of single-qubit rotations acting on qubit $i$ for $i=0,\ldots,n-1$,
which are followed by $n$ CZ gates with control qubit $i$ and target qubit $(i+1) \pmod{n}$
to entangle the quantum states.
Furthermore, input the principal components through single-qubit rotation gates' angles again, namely data re-uploading,
to improve the visualization performance. 
We optimize parameters $\bm{\alpha}_l$ in Eq.~\eqref{eqn:kernel_tsne} with Adam \cite{kingma2014adam}. 
The perplexity is set to $30$ and 
the initial values of $\bm{\alpha}$ are taken from uniform random values between $0$ and $1$. 
We show the resulting low-dimensional data in Fig.~\ref{fig:vis2d}. 
The left figure visualizes the training data, 
and the right one visualizes the test data. 
We see that the present method can nicely form the clusters by the data belonging to each class representing digits.

Next, we quantitatively evaluate the visualization performance of our proposed method.
Specifically, we investigate it by the classification accuracy of low-dimensional data generated by quantum and classical methods. 
If the visualization works well, the low-dimensional data form clusters according to the belonging classes, 
and, in addition, the clusters are separated from the others. 
In such a case, a low-dimensional point is surrounded by points belonging to the same class;
therefore, $k$-nearest neighbor algorithm is effective for evaluating the visualization performance. 
In Ref.~\cite{van2009learning,mcinnes2018umap}, the visualization performance is evaluated 
by the accuracy of $k$-nearest neighbor algorithm of low-dimensional data points. 
In our numerical experiments, we classify the low-dimensional data using $k$-nearest neighbor algorithm
and compare $5$-fold cross-validation accuracy with four conventional models. 
Specifically, the models to be compared are a kernel t-SNE with Gaussian kernels (t-sne(gauss)), 
a parametric t-SNE with a quantum neural network (t-sne(qnn)), 
a parametric t-SNE with a neural network with a single-hidden layer (t-sne(nn1)), and that with three hidden layers (t-sne(nn3)).
We describe their details in Appendix~\ref{sec_appendix:model_detail}.

The accuracies are shown in Table~\ref{tbl:knn}.
The accuracy of t-sne(qnn) is the worst for all values of $k$. 
It is because we could not use many parameters 
due to the long optimization time it causes.
From Table~\ref{tbl:knn}, our model achieves almost the same accuracy with t-sne(gauss)  
but results in a lower accuracy than t-sne(nn1) and t-sne(nn3).
While the present method could not beat the methods based on neural networks, we believe that achieving the comparable performance with respect to the gaussian kernel which can be seen as a basic traditional technique is an encouraging result; in fact, the state-of-the-art result in recognizing hand-written digits using quantum machine learning techniques is at this level \cite{haug2023quantum}.
This showcases the effectiveness of our approach.

\begin{table*}
\centering
\caption{The accuracy of $5$-fold cross-validation by $k$-nearest neighbor algorithm}
\label{tbl:knn}
\begin{tabular}{ |l||c|c|c| }
\hline
                         & $k=1$ & $k=10$ & $k=20$ \\ \hline \hline 
 [proposed method] quantum kernel & $0.792$ & $0.824$ & $0.825$ \\ \hline 
 [t-sne(gauss)] classical gaussian kernel & $0.788$ & $0.816$ & $0.814$ \\ \hline 
 [t-sne(nn1)] parametric t-SNE & $0.787$ & $0.836$ & $0.838$ \\ \hline
 [t-sne(nn3)] parametric t-SNE & $0.870$ & $0.884$ & $0.883$ \\ \hline
 [t-sne(qnn)] quantum neural network  & $0.683$ & $0.728$ & $0.722$ \\ \hline
\end{tabular}
\end{table*}

\subsection{Visualizing quantum data} \label{subsec:numerical_experiments_quantum}
\begin{figure*}
    \centering
	\includegraphics[width=0.6\linewidth]{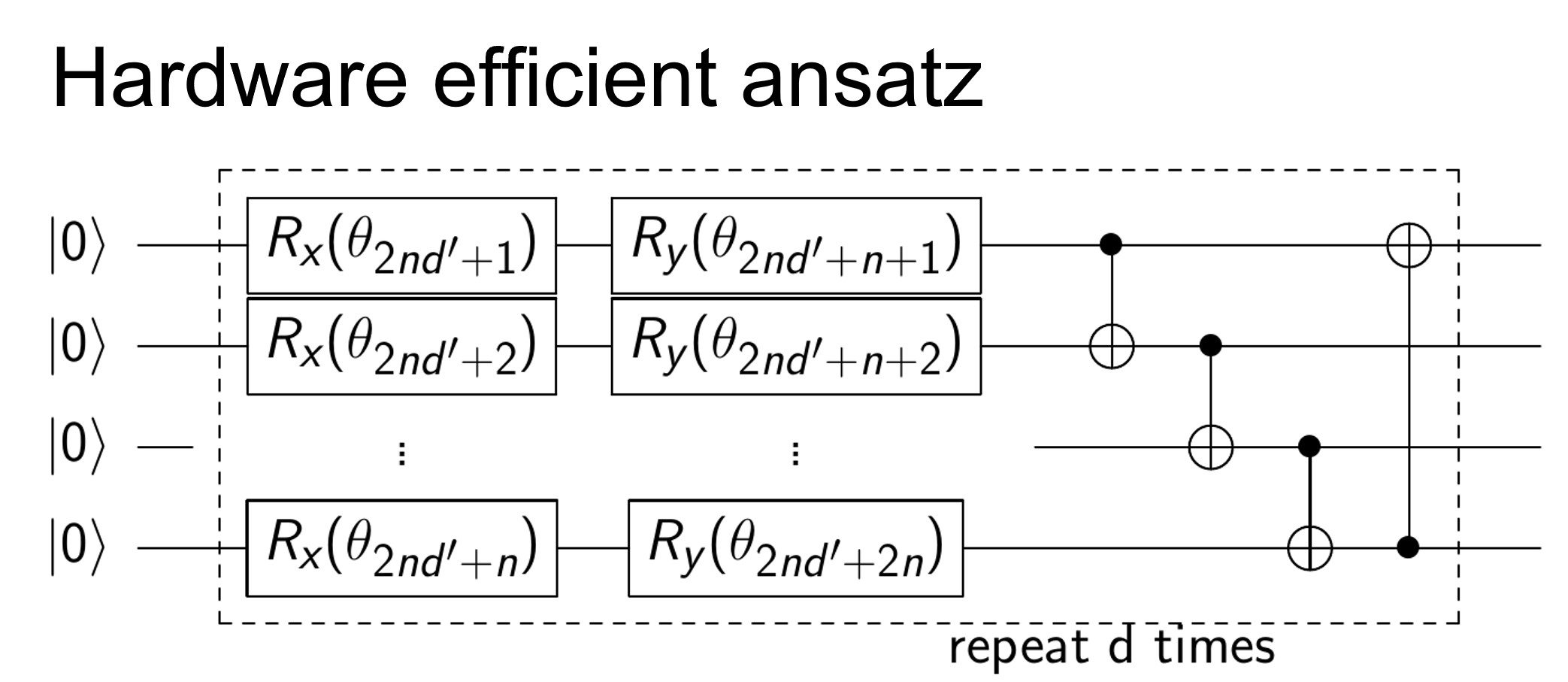}
	\caption{
        This figure shows hardware efficient ansatz, which we use in our numerical experiments. 
        This ansatz consists of single qubit rotation gates and CNOT gates. 
        We set $n=8$ and $d=6$. 
        We repeat the block as $d^\prime=0,1,\ldots,d-1$. 
	}
	\label{fig:hardware_efficient_ansatz}
\end{figure*}

In this section, we present an example of visualizing quantum data. 
More concretely, we visualize quantum states generated within optimization processes of VQE 
applied to transverse field Ising model: 
\begin{equation*}
    H=J\sum_{i=1}^{n-1} Z_i Z_{i+1} + h\sum_{i=1}^n X_i, 
\end{equation*}
where we set $n=8, J=-1.0, h=-0.75$.
The task of the VQE here is to find the ground state of $H$ which we denote by $\ket{\psi_\text{g}}$.
We use the so-called hardware-efficient type ansatz \cite{kandala2017hardware,nakaji2021expressibility}, which is shown in Fig.~\ref{fig:hardware_efficient_ansatz}.
We set the depth $d=6$ and use three different initial parameters drawn from a uniform distribution between $0$ and $2\pi$. 
The parameters are optimized by BFGS algorithm 
implemented in SciPy \cite{2020SciPy-NMeth} with $100$ iterations. 
We visualize the quantum state at each iteration of the optimization by our kernel t-SNE method.
The optimization of parameters in Eq.~\eqref{eqn:kernel_tsne} is performed by Adam \cite{kingma2014adam} setting the perplexity to $10$ and the initial values of $\bm{\alpha}$ to uniform random values between $0$ and $0.1$. 

\begin{figure*}
    \centering
	\includegraphics[width=0.8\linewidth]{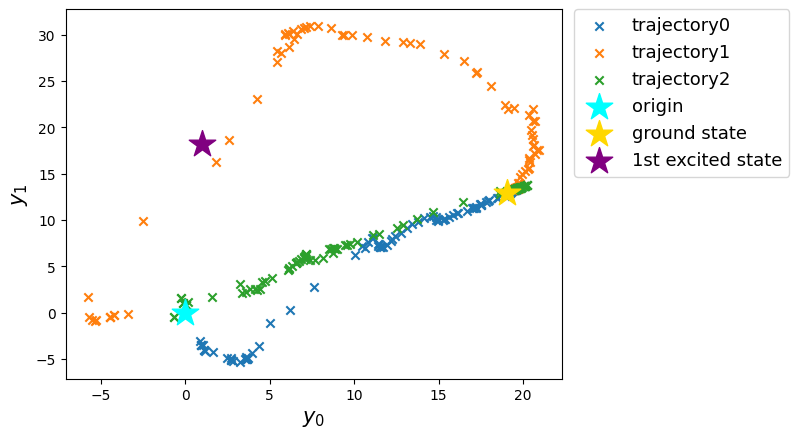}
	\caption{ 
        This figure shows the optimization process of a variational quantum eigensolver 
        to find the ground states of transverse field Ising model. 
        Here, we use hardware efficient ansatz to find the ground state, shown in Figs.~\ref{fig:hardware_efficient_ansatz}. 
        We visualize three trajectories that start from different initial values and label them as trajectory$i$.
	}
	\label{fig:vis_qdata}
\end{figure*}

We show the visualization result in Fig.~\ref{fig:vis_qdata}.
The initial states exist around the origin of the two-dimensional plane
because Eq.~\eqref{eqn:kernel_tsne} returns values that are close to zeros when a quantum state has low similarity to all other quantum states in the data unless $\alpha_l$ is large.
They move away from the origin as the optimizations proceed. 
The trajectories of hardware efficient ansatz converge to the same final state $\ket{\psi_\text{g}}$.
The trajectories of trajectory0 and trajectory2 first diverge but then 
they meet with each other after $30$ iterations.
Their fidelities are actually larger than $0.90$ in this range.
The trajectory of trajectory1, in contrast to the above two, passes nearby the $1$st excited state. 
Making this kind of observation has been impossible with the existing works. 
For instance, our previous approach \cite{kawase2022parametric} demands too much computational resource.
Also, the approach proposed in \cite{rudolph2021orqviz} only looks at the parameters of the circuit and does not visualize quantum states themselves.
With further experiments of this kind, we might be able to extract useful information in characterizing the optimization process of variational quantum algorithms, which may then lead to improved optimizers and ansatz designs.

\section{\label{sec:conclusion}Conclusion}
In this paper, we have proposed a data visualization method using quantum kernels. 
Through numerical experiments, we confirmed that our proposed method could visualize both classical and quantum data. 

For classical data, we visualized a hand-written digits dataset and applied the $k$-nearest neighbor algorithm to low-dimensional data to validate our proposed method. 
The accuracy of our proposed method is almost the same as that of a kernel t-SNE with Gaussian kernels 
but lower than that of a parametric t-SNE with a neural network. 
Note that this result is highly dependent on the choice of quantum kernel employed in the method. 
Because the quantum kernel used in Sec.~\ref{subsec:numerical_experiments_classical} may not be the best choice, 
we might be able to improve the visualization performance by using a data re-uploading circuit \cite{perez2020data,lloyd2020quantum} or optimizing the circuit architecture.
For quantum data, we visualized the optimization process of the VQE. 
We expect this kind of visualization to be helpful in improving initialization and optimization strategies for the VQE.

One interesting future application of our visualization method is to perform quantum architecture search. 
We might be able to construct better quantum circuits more intuitively by visualizing the quantum states that are gerenrated by different architechtures. 
We believe that our visualization method contributes to the development of quantum algorithms by providing means to gain insight about the design of quantum circuits and optimization strategies.

\begin{acknowledgments}
K.M. is supported by JST PRESTO Grant No.~JPMJPR2019 and JSPS KAKENHI Grant No.~20K22330.
This work is supported by MEXT Quantum Leap Flagship Program (MEXTQLEAP) Grant No. JPMXS0118067394 and JPMXS0120319794.
We also acknowledge support from JSTCOI-NEXT program Grant No.~JPMJPF2014.
\end{acknowledgments}

\appendix
\section{\label{sec_appendix:model_detail}The models used in numerical experiments}
\begin{figure*}
    \centering
	\includegraphics[width=0.40\linewidth]{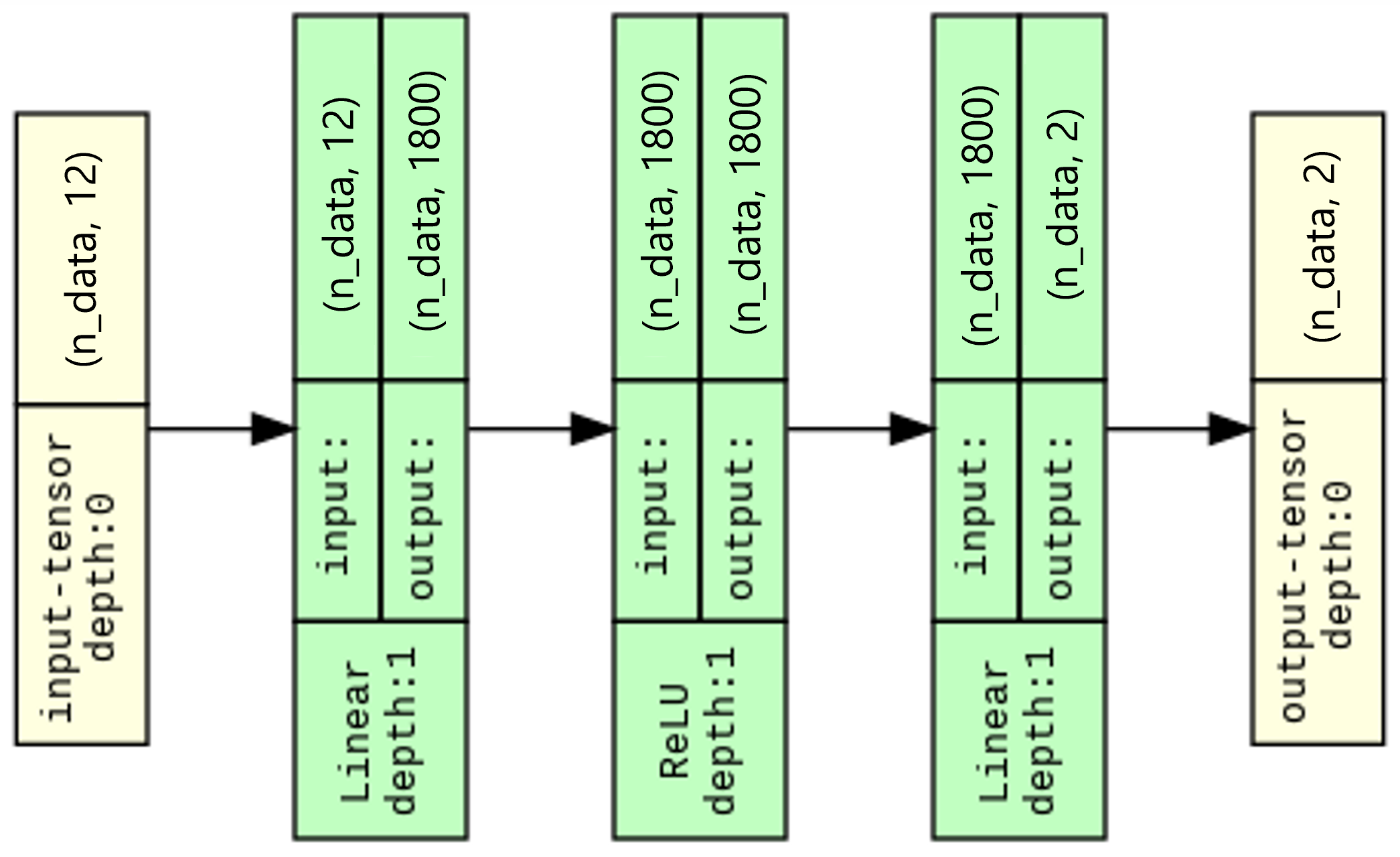}
	\caption{
        This figure shows the architecture of the single-hidden layer neural network
        used in Sec.~\ref{subsec:numerical_experiments_classical}.
        The model consists of an input layer, a hidden layer, the activation function of ReLU, 
        and an output layer. 
	}
	\label{fig:NN1_arch}
\end{figure*}

\begin{figure*}
    \centering
	\includegraphics[width=0.95\linewidth]{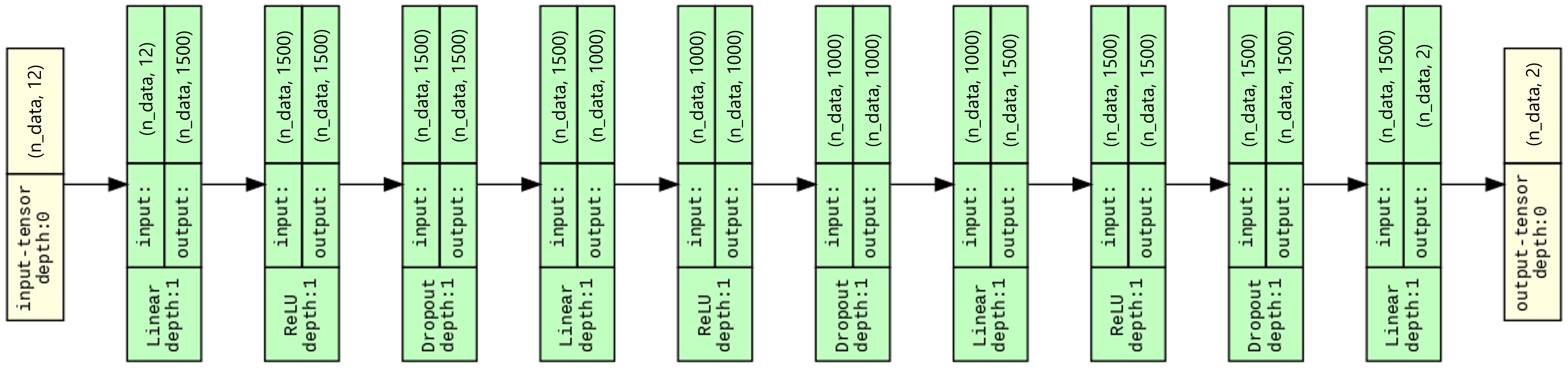}
	\caption{
        This figure shows the architecture of the three-hidden layer neural network 
        used in Sec.~\ref{subsec:numerical_experiments_classical}. 
        The model consists of an input layer, hidden layers, the activation function of ReLU, 
        dropout layers with the rate $0.25$, and an output layer. 
	}
	\label{fig:NN3_arch}
\end{figure*}

\begin{figure*}
    \centering
        \includegraphics[width=0.95\linewidth]{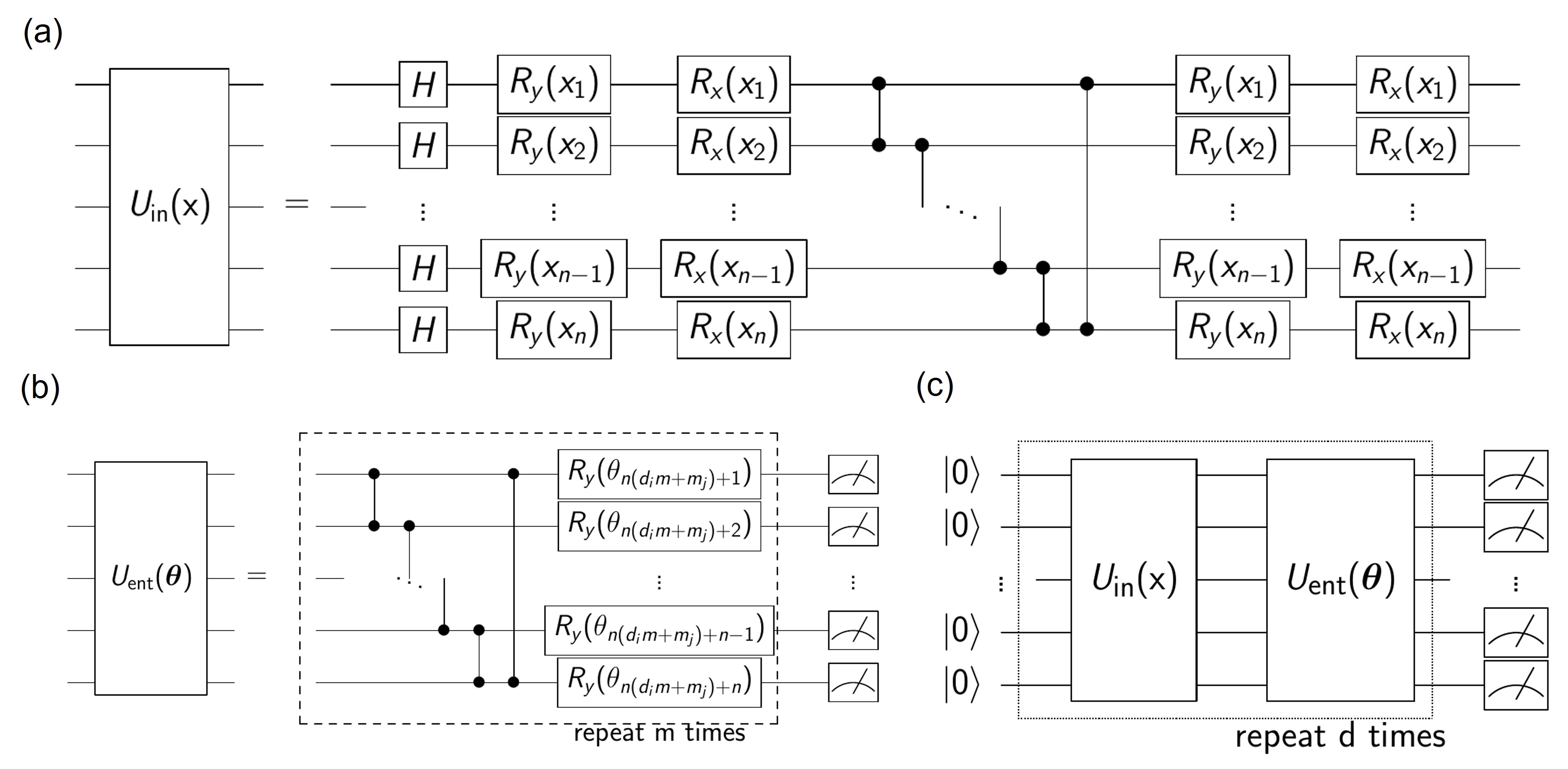}
	\caption{
        This figure shows the architecture of the quantum neural network 
        used in Sec.~\ref{subsec:numerical_experiments_classical}. 
        We encode data by $U_\text{in}(\bm{x})$ as in Fig.~\ref{fig:qnn_arch}(a), 
        and transform them by repeating $U_\text{ent}(\bm{\theta})$ $m=4$ times, consisting of CZ and $R_y$ gates. 
        The angles in the $R_y$ gates in $U_\text{ent}(\bm{\theta})$ are parameters to be optimized. 
        We alternatively repeat $U_\text{in}(\bm{x})$ and $U_\text{ent}(\bm{\theta})$ $d=3$ times. 
	}
	\label{fig:qnn_arch}
\end{figure*}

Here, we explain our models used in numerical experiments. 
Since we described our proposed method in Sec.~\ref{subsec:alg_quantum_kernel_tsne}, 
we explain the other models below. 
\begin{description}
    \item [Kernel t-SNE with Gaussian kernels] 
    As we explained in Sec.~\ref{subsec:kernel_tsne}, 
    the kernel t-SNE is a method of mapping high-dimensional data to low-dimensional space
    using kernel functions 
    so that the similarities between low-dimensional data preserve those between high-dimensional data. 
    Specifically, we adopt the kernel function in Eq.~\eqref{eqn:kernel_tsne} 
    as a Gaussian kernel $k(\bm{x_l},\bm{x})=\exp{(-\|\bm{x_l}-\bm{x}\|^2/2)}$.
    Then, we optimize the $\bm{\alpha}_l$ to minimize the cost function Eq.~\eqref{eqn:costf}. 
    \item [Parametric t-SNE] 
    As we explained in Sec.~\ref{subsec:parametric_tsne}, 
    parametric t-SNE with a neural network is a method of mapping high-dimensional data 
    to a low-dimensional space using a neural network
    in such a way that the similarities between low-dimensional data are almost equal to those between high-dimensional data. 
    First, we explain the architecture of a single-hidden layer neural network used in the numerical experiments. 
    As shown in Fig.~\ref{fig:NN1_arch}, 
    the architecture consists of an input layer, a hidden layer, the activation function of ReLU, and an output layer. 
    We set the number of nodes in the hidden layer to $1800$, which is approximately the same as the number of data.
    Next, we explain the architecture of a three-hidden layer neural network used in the numerical experiments. 
    As shown in Fig.~\ref{fig:NN3_arch}, 
    the architecture consists of an input layer, hidden layers with $1500$, $1000$, and $1500$ nodes 
    with the activation function of ReLU and dropout layers with the rate of $0.25$ for each hidden layer, and then an output layer. 
    Note that 
    the input data we used is that we reduce the dimensions of the given data to $12$ by PCA and standardize each feature of them.
    \item [Parametric t-SNE with a quantum neural network] 
As we explained in Sec.~\ref{subsec:qnn_parametric_tsne}, 
    parametric t-SNE with a quantum neural network is a method of mapping high-dimensional data 
    to low-dimensional space using a quantum neural network
    so that the similarities between low-dimensional data preserve those between high-dimensional data \cite{kawase2022parametric}. 
    As shown in Fig.~\ref{fig:qnn_arch}, 
    the architecture of the quantum neural network consists of 
    data encoding by the circuit described in Fig.~\ref{fig:qc_reupload}
    and unitary transformation including parameterized quantum gates.
    Specifically, we repeat a circuit block consisting of $U_{\text{in}}$ and $U_{\text{ent}}$, alternatively $d=3$ times, where $U_{\text{in}}$ and $U_{\text{ent}}$ are described in Fig.~\ref{fig:qnn_arch}. 
    We input data through $U_{\text{in}}$, consisting of $R_x$, $R_y$, and CZ gates, and transform them by repeating $U_{\text{ent}}$ $m=4$ times, consisting of CZ and $R_y$ gates. 
    Note that we normalize each feature of the data to fit within $[0.0,0.5]$ so that the principal components are ranged within $[-1.0,1.0]$. 
    Then, we optimize the angles $\bm{\theta}$ in $R_y$ gates  in $U_{\text{ent}}$. 
    Let us denote the map of the quantum neural network as $U(\bm{x}_i, \bm{\theta})$, 
    the low-dimensional data are described by
    \begin{align*}
        y_{i,0}(\bm{x}_i, \bm{\theta}) &= \sum_{j=1}^{n/2} \beta_j \bra{0}U^\dagger(\bm{x}_i,\bm{\theta}) X_j U(\bm{x}_i,\bm{\theta})\ket{0}, \\
        y_{i,1}(\bm{x}
_i, \bm{\theta}) &= \sum_{j=n/2+1}^{n} \beta_j \bra{0}U^\dagger(\bm{x}_i,\bm{\theta}) X_j U(\bm{x}_i,\bm{\theta})\ket{0}, 
    \end{align*}
    where we optimize both $\beta_j$ and $\bm{\theta}$ to minimize the cost function. 
\end{description}

\bibliography{apssamp}

\end{document}